\documentclass[twocolumn,showpacs,amsmath,amssymb]{revtex4}
\usepackage{amsmath}

\usepackage{graphicx}
\usepackage{amsfonts}% Include figure files
\usepackage{dcolumn}% Align table columns on decimal point
\usepackage{bm}% bold math

\begin{document}

\title{Optimal modularity for nucleation in network-organized Ising model}

\author{Hanshuang Chen}

\author{Zhonghuai Hou}\email{hzhlj@ustc.edu.cn}

\affiliation{Hefei National Laboratory for Physical Sciences at
 Microscales \& Department of Chemical Physics, University of
 Science and Technology of China, Hefei, 230026, China}

\date{\today}

\begin{abstract}
We study nucleation dynamics of Ising model in a topology that
consists of two coupled random networks, thereby mimicking the
modular structure observed in real-world networks. By introducing a
variant of a recently developed forward flux sampling method, we
efficiently calculate the rate and elucidate the pathway for
nucleation process. It is found that as the network modularity
becomes worse the nucleation undergoes a transition from two-step to
one-step process. Interestingly, the nucleation rate shows a
nonmonotonic dependency on the modularity, in which a maximal
nucleation rate occurs at a moderate level of modularity. A simple
mean field analysis is proposed to qualitatively illustrate the
simulation results.
\end{abstract}
\pacs{89.75.Hc, 64.60.Q-, 05.50.+q} \maketitle

\section{Introduction} \label{sec1}

In the last decade, critical phenomena in complex networks have
received an enormous amount of attention in the field of statistical
physics and many other disciplines (see, for example, a recent
review \cite{RMP08001275}). Extensive research interests have
focused on the onset of critical behaviors in diverse network
topology, which included a wide range of issues: percolation
phenomenon \cite{PRL00004626, PRL00005468, PRL02208701,
PRE02036113}, epidemic thresholds
\cite{PRL01003200,PhysRevLett.105.218701}, order-disorder
transitions \cite{PHA02000260, PLA02000166, PRE02016104,
EPB02000191,PRE04067109}, synchronization \cite{PRL03014101,
PRL07034101}, self-organized criticality \cite{PRL03148701,
PRE002065102}, nonequilibrium pattern formation \cite{Nat10000544},
etc. However, there is much less attention paid to the
dynamics/kinetics of phase transition itself in complex network,
such as nucleation in a first-order phase transition.

Nucleation is a fluctuation-driven process that initiates the decay
of a metastable state into a more stable one \cite{Kashchiev2000}. A
first-order phase transition usually involves the nucleation and
growth of a new phase. Many important phenomena in nature, including
crystallization \cite{JCP97003634}, glass formation
\cite{PRE98005707}, and protein folding \cite{PNAS9510869}, etc.,
are associated with nucleation. Despite its apparent importance,
many aspects of nucleation process are still unclear and deserve
more investigations. The Ising model, which is a paradigm for many
phenomena in statistical physics, has been widely used to study the
nucleation process. Despite its simplicity, Ising model has made
important contributions to the understanding of nucleation phenomena
in equilibrium systems and is likely to yield important insights
also for nonequilibrium systems. In two-dimensional lattices, for
instance, shear can enhance the nucleation rate and the rate peaks
at an intermediate shear rate \cite{JCP08134704}, a single impurity
may considerably enhance the nucleation rate \cite{JPC06004985}, and
the existence of a pore may lead to two-stage nucleation and the
overall nucleation rate can reach a maximum level at an intermediate
pore size \cite{PRL06065701}. Nucleation pathway of Ising model in
three-dimensional lattice has also been studied using transition
path sampling approach \cite{JPC0419681}. In addition, Ising model
has been frequently used to test the validity of classical
nucleation theory (CNT) \cite{EPJ98000571,JCP99006932,JCP00001976,
PRE05031601,PRE10030601}. However, all these studies are limited to
regular lattices in Euclidean space. Since many real systems can be
properly modeled by network-organized structure, it is thus natural
to ask how the topology of a networked system affects the nucleation
process of Ising model.

In a recent work \cite{arXiv:1008.0704}, we have studied nucleation
dynamics on scale-free networks, in which we found that nucleation
starts from, on average, nodes with more lower degrees, and the rate
for nucleation decreases exponentially with network size and the
size of critical nucleus increase linearly with network size,
implying that nucleation is relevant only for a finite-size network.
Herein, we want to study nucleation dynamics of Ising model in a
modular network composed of two coupled random networks. It is known
that many real-world networks, as diverse as from social networks to
biological networks, have found to exhibit modularity structures
\cite{PNAS06008577,PhysRep10000075}, that is, links within modules
are much more denser than those between modules. Many previous
studies have revealed that such modular structures have a
significant impact on the dynamics taking place on the networks,
such as synchronization
\cite{PhysRevLett.96.114102,PhysRevLett.101.168701}, neural
excitability \cite{PhysRevLett.97.238103}, spreading dynamics
\cite{EPL05000315,PhysRevE.73.035103}, opinion formation
\cite{JSM0708026,PhysRevE.75.030101}, and Ising phase transition
\cite{EPL0968006,PRE09025101,PRE09031110}. In particular, for
majority model \cite{JSM0708026} and Ising model
\cite{EPL0968006,PRE09025101,PRE09031110} in modular networks, it
has been shown that there exists a region in a discontinuous
transition where modular order phase and global order phase coexist.
However, these studies mainly focused on phase diagrams in
parameters space, and did not make detailed investigation for the
transition process from a phase to another that may undergo a
nucleation process.

Since nucleation is an activated process that occurs extremely slow,
brute-force simulation is prohibitively expensive. To overcome this
difficulty, we will use a variant of a recently developed simulation
method, forward flux sampling (FFS) \cite{PRL05018104}. This method
allows us to calculate nucleation rate and determine the properties
of ensemble toward nucleation pathways. We found that as the degree
of network modularity decreases nucleation goes through a transition
from two-step to one-step process, and the rate exhibits a maximum
at an intermediate degree of modularity. Free energy profiles for
different modularity obtained by umbrella sampling (US)
\cite{JCP92000015} and a simple mean-field (MF) analysis help us
understand the FFS results.

This paper is organized as follows. In Sec.\ref{sec2}, we describe
the details of our model and the simulation method applied to this
system. In Sec.\ref{sec3}, we present the results for the nucleation
rate and pathway in modular networks. In Sec.\ref{sec4}, a simple
mean field analysis is used to qualitatively illustrate the
simulation results. At last, discussion and main conclusions are
addressed in Sec.\ref{sec5}.

\section{Model and Simulation details} \label{sec2}

Consider a network consisting of $N$ nodes arranged into two modules
with $N_1$ and $N_2$ nodes. For simplicity, we only consider the
case of $N_1=N_2=N/2$ throughout this paper. The connection
probability between a pair of nodes belonging to the same module is
$\rho_i$, while that for nodes belonging to different modules is
$\rho_o$. The parameter $\sigma=\frac{{\rho _o }} {{\rho _i }} \in
\left[ {0,1} \right]$, defined as the ratio of inter- to
intra-modular connectivity, measures the degree of modularity. The
higher degree of modularity of a network is, the smaller value of
$\sigma$ is. As $\sigma \to 0$, the network becomes two isolated
clusters, while as $\sigma \to 1$, the network approaches a
Erd\"os--R\'enyi (ER) random network. When $\sigma$ is varied the
total number of links of the network is kept unchanged,
${\frac{{N\left\langle k \right\rangle }} {2}}$, where $\left\langle
k \right\rangle$ is the average degree. This restriction leads to
$\rho _i  = \frac{{2\left\langle k \right\rangle }} {{N\left( {1 +
\sigma } \right)}}$ and $\rho _o  = \frac{{2\left\langle k
\right\rangle \sigma }} {{N\left( {1 + \sigma } \right)}}$. Each
node is endowed with an Ising spin variable $s_i$ that can be either
$+1$ (up) or $-1$ (down). The Hamiltonian of the system is given by
\begin{equation}
H=-J\sum\nolimits_{i < j}{a_{ij}s_i s_j}-h\sum\limits_i s_i,
\label{eq1}
\end{equation}
where $J(>0)$ is the coupling constant and $h$ is the external
magnetic field. The elements of the adjacency matrix of the network
take $a_{ij}=1$ if nodes $i$ and $j$ are connected and $a_{ij}=0$
otherwise.

Our simulation is performed by Metropolis spin-flip dynamics
\cite{Lan2000}, in which we attempt to flip each spin once, on
average, during each Monte Carlo (MC) cycle. In each attempt, a
randomly chosen spin is flipped with the probability $\min(1,e^{-
\beta \Delta E})$, where $\beta=1/(k_B T)$ and $k_B$ is the
Boltzmann constant and $T$ is the temperature, and $\Delta E$ is the
energy difference due to the flipping process. Here, we set $J=1$,
$h>0$, $T<T_c$ ($T_c$ is the critical temperature), and start with a
metastable state in which $s_i=-1$ for most of the spins. The system
will stay in that state for a significantly long time before
undergoing a nucleation transition to a more stable state with most
spins pointing up. We are interested in the pathway and rate for
this nucleation process.

FFS method has been used to calculate rate constants and transition
paths for rare events in equilibrium and nonequilibrium systems
\cite{PRL05018104, PRL06065701, JCP08134704, JPC06004985,
JCP07114109, JCP06024102}. This method uses a series of interfaces
in phase space between the initial and final states to force the
system from the initial state $A$ to the final state $B$ in a
ratchet-like manner. Before the simulation begin, an order parameter
$\lambda$ is first defined, such that the system is in state $A$ if
$\lambda<\lambda_0$ and it is in state $B$ if $\lambda>\lambda_M$. A
series of nonintersecting interfaces $\lambda_i$ ($0<i<M$) lie
between states $A$ and $B$, such that any path from $A$ to $B$ must
cross each interface without reaching $\lambda_{i+1}$ before
$\lambda_i$. The algorithm first runs a long-time simulation which
gives an estimate of the flux $\bar \Phi_{A,0}$ escaping from the
basin of $A$ and generates a collection of configurations
corresponding to crossings of interface $\lambda_0$. The next step
is to choose a configuration from this collection at random and use
it to initiate a trial run which is continued until it either
reaches $\lambda_1$ or returns to $\lambda_0$. If $\lambda_1$ is
reached, store the configuration of the end point of the trial run.
Repeat this step, each time choosing a random starting configuration
from the collection at $\lambda_0$. The fraction of successful trial
runs gives an estimate of of the probability of reaching $\lambda_1$
without going back into A, $P\left( {\lambda_1 |\lambda_0} \right)$.
This process is repeated, step by step, until $\lambda_M$ is
reached, giving the probabilities $P\left( {\lambda_{i+1}
|\lambda_i} \right)$ ($i=1, \cdots,M-1$). Finally, we get the
transition rate $R$ from $A$ to $B$, which is the product of the
flux $\bar \Phi_{A,0}$ and the probability $P\left( {\lambda_{M}
|\lambda_0} \right) = \prod\nolimits_{i=0}^{M-1}{P\left( {\lambda_{i
+ 1} |\lambda_i} \right)}$ of reaching $\lambda_M$ from $\lambda_0$
without going into $A$. The detailed descriptions of FFS method see
Ref.\cite{JPH09463102}.

However, conventional FFS method will become very inefficient if one
intermediate metastable state exists between initial state and final
state, as a two-step nucleation process demonstrated in
Fig.\ref{fig1}. This is because that sampling paths will be trapped
in these long-lived metastable states so that they hardly return to
initial state. To solve this problem, we will perform instead
two-step samplings from initial down-spin state to intermediate
metastable state, and then to final up-spin state, giving the
two-step rates, $R_1$ and $R_2$, respectively. Since the total mean
time for nucleation is simply the sum of the mean time of the
two-step process, the total rate can be expressed as $R = \left(
{R_1^{ - 1} + R_2^{ - 1} } \right)^{ - 1}$. To determine the
location of the intermediate state, during FFS we monitor the
sampling time for the probability $P\left( {\lambda_{i+1}
|\lambda_i} \right)$ between two neighboring interfaces. If the
sampling time spent on between interfaces $i$ and $i+1$ is much more
than its previous step and the probability $P\left( {\lambda_{i+1}
|\lambda_i} \right)$ is nearly one, we consider the $i$th interface
as the location of the intermediate state. If such conditions do not
meet during the whole sampling, the intermediate metastable state
does not exist, meaning that nucleation is a one-step process.  Note
that the method is straightforward to generalize to a multi-step
nucleation process.

Here, we define the order parameter $\lambda$ as the total number of
up spins in the networks. The spacing between interfaces is fixed at
$3$ up spins, but the computed results do not depend on this
spacing. We perform $1000$ trials per interface for each FFS
sampling, from which at least $100$ configurations are saved at each
interface in order to investigate the statistical properties of an
ensemble of reactive pathways to nucleation. The results are
obtained by averaging over $10$ independent FFS samplings and $50$
different network realizations.

\section{Results} \label{sec3}

\begin{figure}
\centerline{\includegraphics*[width=0.9\columnwidth]{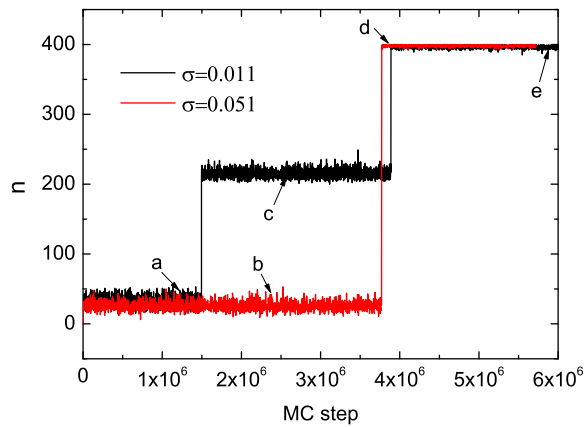}}
\caption{ (Color online) Typical time evolutions of the number of up
spins $\lambda$. It is shown that the system undergoes a two-step
nucleation process for $\sigma=0.011$ and a one-step nucleation
process for $\sigma=0.051$. The representative network
configurations at different moments indicated by arrows are shown in
Fig.\ref{fig2}. Other parameters are $N=400$, $\left\langle k
\right\rangle =6$, $T=2.0$, and $h=1.2$. \label{fig1}}
\end{figure}

\begin{figure}
\centerline{\includegraphics*[width=1.0\columnwidth]{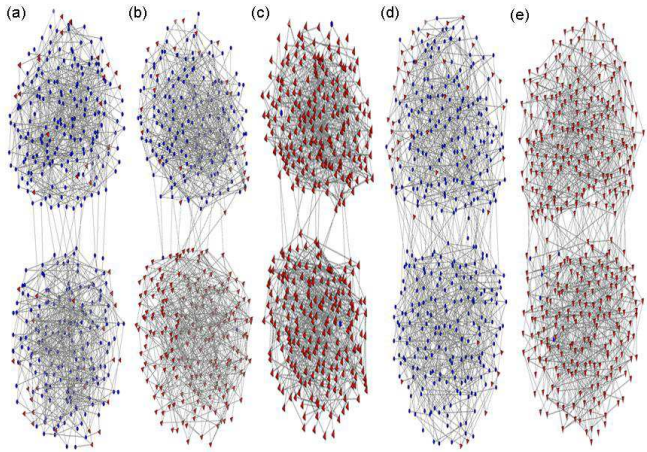}}
\caption{ (Color online) Five representative network configurations
at different moments indicated in Fig.\ref{fig1}, where down-spin
nodes and up-spin nodes are denoted by blue circles and red
triangles, respectively. (a)-(c) correspond to the case of
$\sigma=0.011$ and (d)-(e) correspond to the case of $\sigma=0.051$.
\label{fig2}}
\end{figure}

To begin with, in Fig.\ref{fig1} we exhibit typical time evolutions
of the number of up spins $\lambda$ corresponding to two different
values of network modularity $\sigma=0.011$ and $\sigma=0.051$ via
brute-force simulations, with relevant parameters $N=400$,
$\left\langle k \right\rangle =6$, $T=2.0$, and $h=1.2$. It is
clearly observed that the system undergoes a two-step nucleation
process for $\sigma=0.011$ and a one-step nucleation process for
$\sigma=0.051$. We also plot several representative configurations
in Fig.\ref{fig2}, corresponding to different phases of the system,
respectively. Before the nucleation happening, the system lies in a
metastable state, where most of the nodes are in down-spin state
(indicated by blue circles), as shown in Fig.\ref{fig2}(a) and
Fig.\ref{fig2}(d). When the network modularity is very good, the
system enters into an intermediate metastable state via the
first-step nucleation, where nodes in one of modules are in up-spin
state (indicated by red triangles), while nodes in the other module
are still in down-spin state, as shown in Fig.\ref{fig2}(b). When
the network modularity becomes worse, such an intermediate
metastable state disappears so that the nucleation becomes a
one-step process. Finally, the system will enter into the most
stable state, where almost all spins are in up-spin state, as shown
in Fig.\ref{fig2}(c) and Fig.\ref{fig2}(e). Moreover, we note that
that the nucleation process typically takes the order of $10^6$ or
more MC steps  that is computationally costly. Therefore, in what
follows we will give the results obtained by FFS method.

\begin{figure}
\centerline{\includegraphics*[width=0.9\columnwidth]{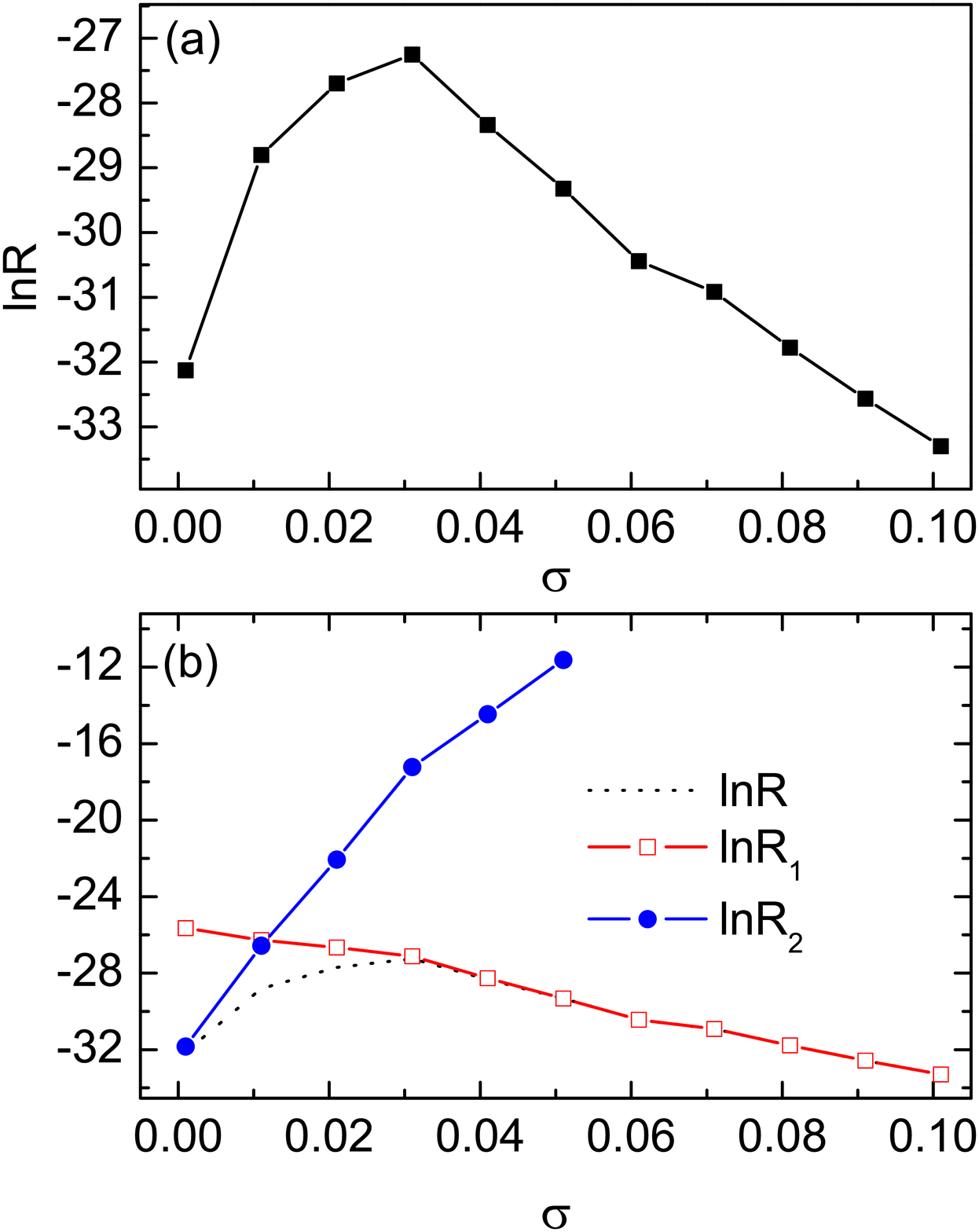}}
\caption{(Color online) (a) The logarithm of nucleation rare $lnR$
as a function of the degree of modularity $\sigma$. (b) $lnR_1$
(squares) and $lnR_2$ (circles) as a function of $\sigma$, and
dotted line indicates the overall rate $lnR$. The parameters are
same as in Fig.\ref{fig1} except for $h=1.0$. \label{fig3}}
\end{figure}

The nucleation rate $R$ as a function of $\sigma$ is plotted in
Fig.\ref{fig3}(a), with relevant parameters being the same as those
in Fig.\ref{fig1} except for $h=1.0$. One can see that as $\sigma$
increases $R$ reaches a maximum $R_c$ at $\sigma \simeq 0.031$ and
then decreases. Obviously, there exists a maximal nucleation rate
that occurs at a moderate degree of network modularity. In
Fig.\ref{fig3}(b), we plot the results of the nucleation rates,
$R_1$ and $R_2$, for two-steps process as functions of $\sigma$. As
$\sigma$ increases, $R_1$ seems to exponentially decreases with
$\sigma$, while $R_2$ increases monotonically until $\sigma=0.051$
is reached. For $\sigma>0.051$, nucleation becomes one-step process
so that $R_2$ can not be well defined and the overall nucleation
rate is only determined by $R_1$. From Fig.\ref{fig3}(b) one can
find that $R_2$ is much lower than $R_1$ when the value of $\sigma$
is relatively small, so that $R$ is dominantly determined by the
second step nucleation. While for $\sigma > 0.031$, $R$ is almost
determined only by the first step nucleation. Thus, there exists a
region $0.001<\sigma<0.031$ where $R$ is determined by both $R_1$
and $R_2$. Note that we have also made extensive simulations for
other parameters such as $h=0.7,1.2$ and $T=1.5,1.8$, and found that
the qualitative features of the above results do not change (results
now shown).

\begin{figure}
\centerline{\includegraphics*[width=0.9\columnwidth]{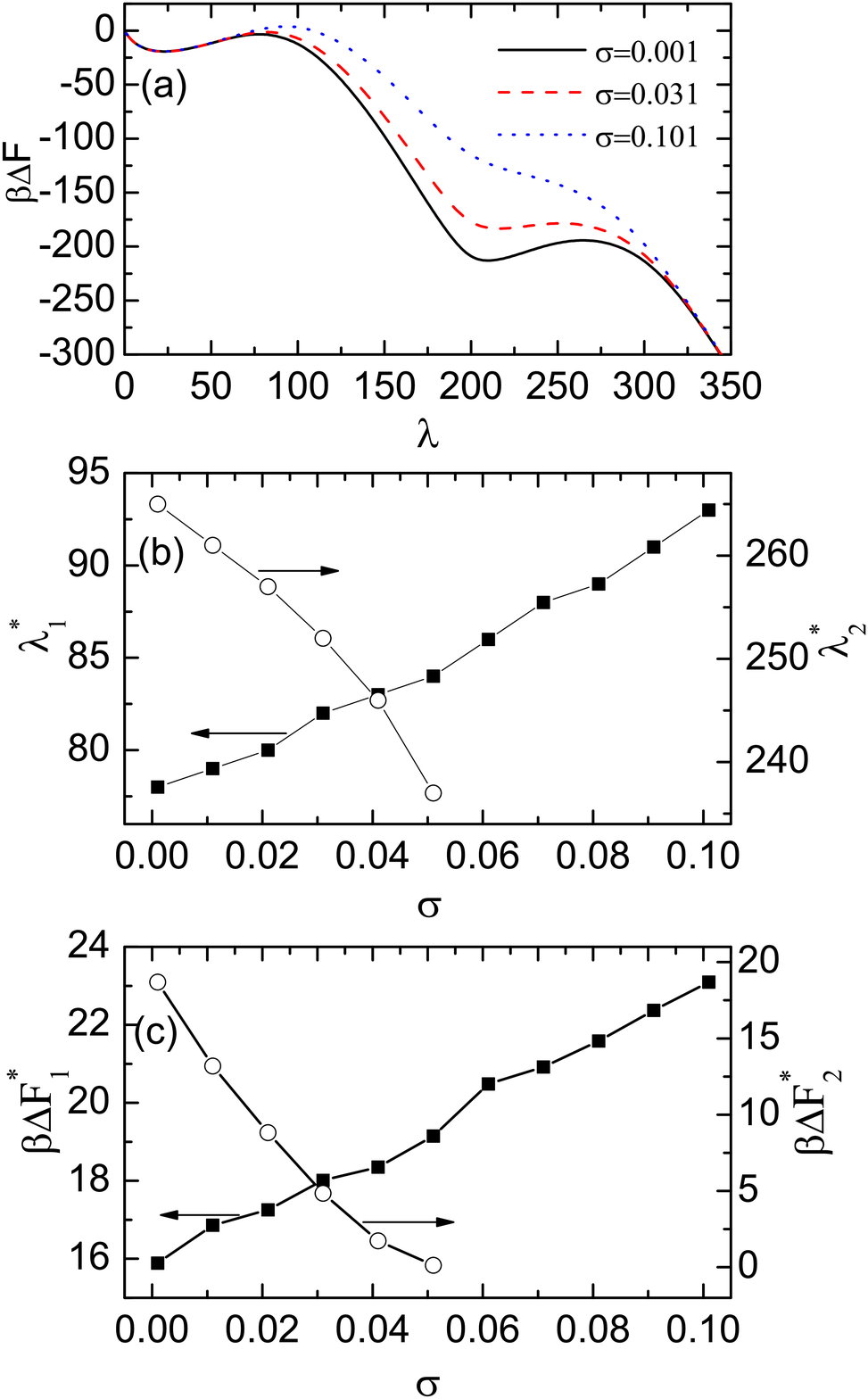}}
\caption{(Color online) (a) Free energy $\Delta F$ as a function of
$\lambda$ for three different $\sigma=0.001, 0.031, 0.101$. For
smaller $\sigma$ two free-energy barriers are clearly observed,
while for larger $\sigma$ the second one vanishes. (b) The size of
critical nucleus $\lambda_1^*$ and $\lambda_2^*$, and (c) the
free-energy barriers $\Delta F_1^*$ and $\Delta F_2^*$, as functions
of $\sigma$. The other parameters are the same as Fig.\ref{fig2}.
\label{fig4}}
\end{figure}

To further understand the above results, we calculate free energy of
the system by US method, in which we have used a bias potential
$0.1k_B T(\lambda-\bar \lambda)^2$, with $\bar \lambda$ being the
center of each window. The free energy $\Delta F$ as a function of
$\lambda$ for three different values of $\sigma$ are depicted in
Fig.\ref{fig4}(a). For $\sigma=0.001$ and $\sigma=0.031$, there are
two free-energy maximums, occurring at the locations of critical
nucleus of $\lambda=\lambda_1^*$ and $\lambda=\lambda_2^*$,
respectively. This picture is consistent with the two-step
nucleation process described before. For a larger $\sigma=0.101$,
just the first free-energy barrier is present, implying that the
nucleation becomes one-step process. With the increment of $\sigma$,
$\lambda_1^*$ moves to a larger value while the value of
$\lambda_2^*$ gets smaller, as shown in Fig.\ref{fig4}(b).
Fig.\ref{fig4}(c) shows that the first free-energy barrier $\Delta
F_1^*$, defined as the difference between the free energy at
$\lambda_1^*$ and the first minimum in free energy ($\lambda=23$),
nearly increases linearly with $\sigma$, while the second
free-energy barrier $\Delta F_2^*$ (definition is similar to that of
$\Delta F_1^*$, and the second minimum in free energy is an
increasing function of $\sigma$, within the range $\lambda \in [78,
93]$) decreases monotonically with $\sigma$ until $\Delta F_2^*$
vanishes at $\sigma>0.051$, which is in agreement with the result of
Fig.\ref{fig3}(b).

\section{Mean field analysis} \label{sec4}

\begin{figure}
\centerline{\includegraphics*[width=0.9\columnwidth]{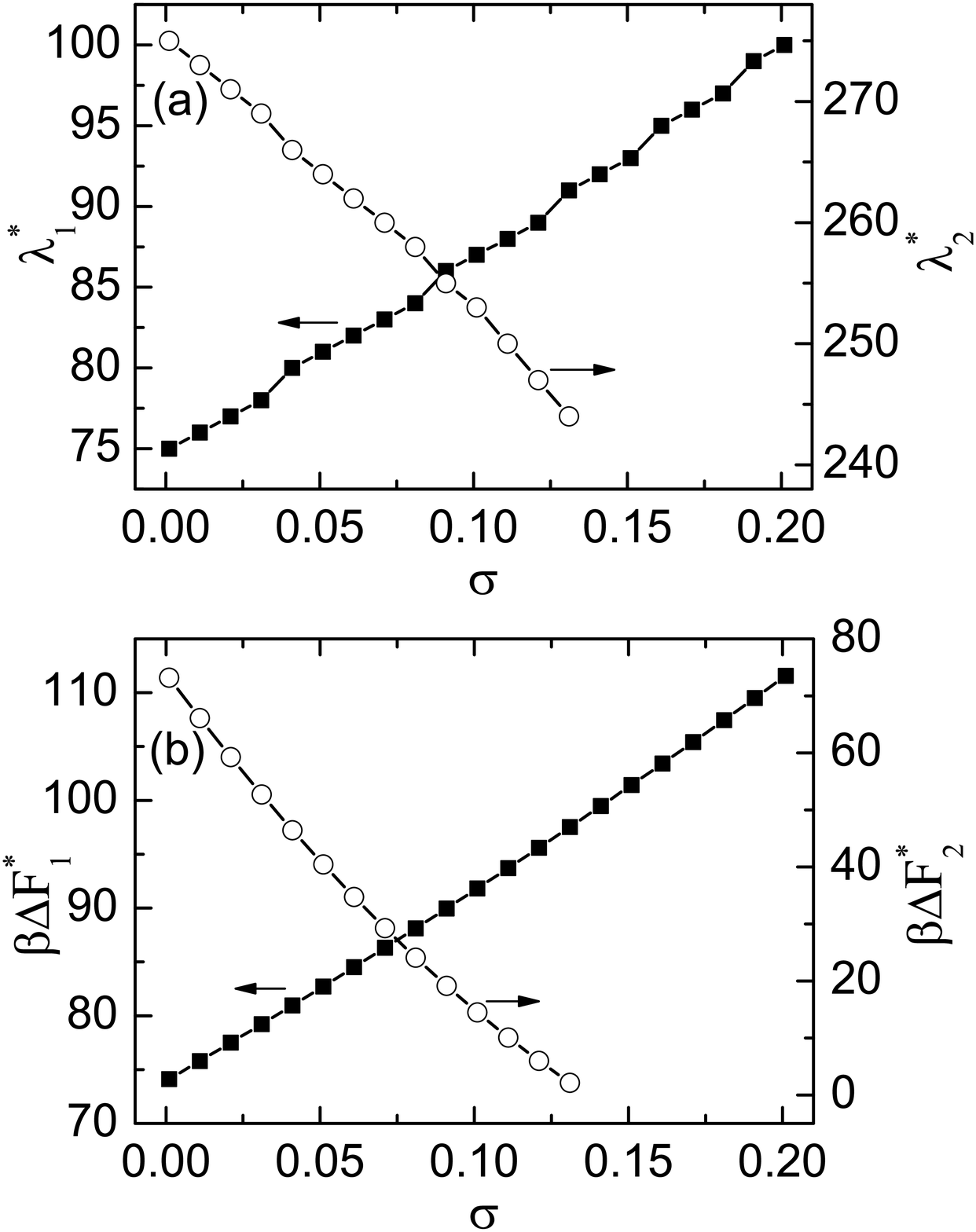}}
\caption{The results of mean field analysis. (a) The size of
critical nucleus $\lambda_1^*$ and $\lambda_2^*$, and (b) the
free-energy barriers $\Delta F_1^*$ and $\Delta F_2^*$, as functions
of $\sigma$. The other parameters are the same as Fig.\ref{fig2}.
\label{fig5}}
\end{figure}

In order to unveil possible mechanism behind the above phenomenon,
we will present an analytical understanding by CNT and simple MF
approximation. Firstly, let us assume, for the first-step
nucleation, that $\lambda$ nodes are in up-spins and the remaining
nodes are in down-spins in one of modules (say module I for
convenience), and all the nodes in the other module (module II) are
in down-spins. The energy change due to the spin-flip of these
$\lambda$ nodes can be expressed as the sum of two parts $\Delta
U_1=-2h \lambda+2JN^{in}_1$, where the first part denotes the energy
loss due to the creation of $\lambda$ up-spins, which favors the
growth of the nucleus, while the second part denotes the energy gain
due to the formation of $N^{in}_1$ new interfacial links between up
and down spins, which disfavors the growth of the nucleus. According
to MF approximation, $N^{in}_1$ can be written as $N^{in}_1=\rho _i
\lambda \left( {\frac{N} {2} - \lambda } \right) + \rho _o \lambda
\frac{N} {2}$, where the first part and the second part arise from
interfacial links inside module I and between modules, respectively.
For the second-step nucleation, we assume that all the nodes in
module I are in up-spins, and $\lambda$ nodes are in up-spins and
the remaining nodes are in down-spins in module II. This process
creates new interfacial links inside module II, and at the same time
removes old interfacial links between module I and module II. Thus,
the energy change for this process is $\Delta
U_2=-2h\lambda+2JN^{in}_2$ where $N^{in}_2= \rho _i \lambda \left(
{\frac{N} {2} - \lambda } \right) - \rho _o \lambda \frac{N} {2}$ is
the net number of the interfacial links. The entropy changes for the
two nucleation processes are both $\Delta S=- \frac{{k_B N}}
{2}\left[ {\frac{{2\lambda }} {N}\ln \left( {\frac{{2\lambda }} {N}}
\right) + \left( {1 - \frac{{2\lambda }} {N}} \right)\ln \left( {1 -
\frac{{2\lambda }} {N}} \right)} \right]$. Then, the changes of free
energy for the two-step processes are $\Delta F_i  = \Delta U_i  -T
\Delta S$ ($i=1,2$). In Fig.\ref{fig5} we give the analytical
results of the critical nucleus and free-energy barriers as
functions of the network modularity. Clearly, the analysis
qualitatively agrees with the simulation results of Fig.\ref{fig4}.
From Fig.\ref{fig5}, one can see that with the increment of $\sigma$
the size of the first critical nucleus and the height of the first
free-energy barrier increase almost linearly, while the size of the
second critical nucleus and the height of the second free-energy
barrier decrease until $\sigma \simeq 0.13$ is reached. This implies
that the analysis also predicts the extinction of the second
nucleation stage, but this prediction obviously overestimates the
transition value of $\sigma$.

\section{Conclusions} \label{sec5}

In conclusion, we have studied nucleation dynamics of Ising model in
modular networks consisted of two random networks. Using FFS method,
we found that as the network modularity gradually becomes worse a
transition occurs from one-step to two-step nucleation process.
Interestingly, the nucleation rate is a nonmonotonic function of the
degree of modularity and a maximal rate exists for an intermediate
level of modularity. Using US method, we obtained free energy
profiles at different network modularity, from which one can see
that two free-energy barriers exist for very good modularity and the
second one vanishes when the network modularity becomes worse. This
picture further confirms the FFS results. Finally, a mean field
analysis is employed to understand the nature of nucleation in
modular networks and the simulation results. Since stochastic
fluctuation and the coexistent of multi-states are ubiquitous in
social and biological systems, our study may shed valuable insights
into fluctuation-driven transition phenomena taking place in
network-organized systems with modular structures.

\begin{acknowledgments}
This work was supported by NSFC (Nos. 20933006, 20873130).
\end{acknowledgments}

%\bibliographystyle{apsrev}
%\bibliography{Nucleation_RN}

\end{document}